\pdfoutput=1

\documentclass[10pt]{wlscirep}

\usepackage[utf8]{inputenc}
\usepackage[T1]{fontenc}
\usepackage{setspace}
\usepackage{siunitx}
\usepackage[version=4]{mhchem}

\newcommand{\figref}[1]{Fig.~{\ref{#1}}}

\newif\ifsubmit
\submittrue

\ifsubmit
    \newcommand{\can}[1]{}
    \newcommand{\py}[1]{}
    \newcommand{\todo}[1]{}
    \newcommand{\tocite}[1]{}
\else
    \definecolor{comments}{rgb}{0.1, 0.66, 0.1}
    \newcommand{\can}[1]{[{\color{comments}CL: #1}]}
    \newcommand{\py}[1]{[{\color{brown}Peiyi: #1}]}
    \newcommand{\todo}[1]{[{\color{red}TODO: #1}]}
    \newcommand{\tocite}[1]{[{\color{red}citation-#1}]}
\fi

\title{Real-time raw signal genomic analysis using fully integrated memristor hardware}

\author[1]{Peiyi He}
\author[1]{Shengbo Wang}
\author[1]{Ruibin Mao}
\author[2]{Sebastian Siegel}
\author[3]{Giacomo Pedretti}
\author[3]{Jim Ignowski}
\author[2,4]{John Paul Strachan}
\author[5,*]{Ruibang Luo}
\author[1,*]{Can Li}

\affil[1]{Department of Electrical and Electronic Engineering, The University of Hong Kong, Hong Kong SAR, China}
\affil[2]{Peter Grünberg Institute (PGI-14), Forschungszentrum J\"ulich GmbH, J\"ulich, Germany}
\affil[3]{Hewlett Packard Labs, Hewlett Packard Enterprise, Milpitas, CA, USA}
\affil[4]{RWTH Aachen University, Aachen, Germany}
\affil[5]{Department of Computer Science, School of Computing and Data Science, The University of Hong Kong, Hong Kong SAR, China}
\affil[*]{e-mail: rbluo@cs.hku.hk, canl@hku.hk}

\begin{abstract}
Advances in third-generation sequencing have enabled portable and real-time genomic sequencing, but real-time data processing remains a bottleneck, hampering on-site genomic analysis due to prohibitive time and energy costs.
These technologies generate a massive amount of noisy analog signals that traditionally require basecalling and digital mapping, both demanding frequent and costly data movement on von Neumann hardware. 
To overcome these challenges, we present a memristor-based hardware-software co-design that processes raw sequencer signals directly in analog memory, effectively combining the separated basecalling and read mapping steps.
Here we demonstrate, for the first time, end-to-end memristor-based genomic analysis in a fully integrated memristor chip. 
By exploiting intrinsic device noise for locality-sensitive hashing and implementing parallel approximate searches in content-addressable memory,
we experimentally showcase on-site applications including infectious disease detection and metagenomic classification. 
Our experimentally-validated analysis confirms the effectiveness of this approach on real-world tasks, achieving a state-of-the-art 97.15\% F1 score in virus raw signal mapping, with 51\textbf{$\times$} speed up and 477\textbf{$\times$} energy saving compared to implementation on a state-of-the-art ASIC.
These results demonstrate that memristor-based in-memory computing provides a viable solution for integration with portable sequencers, enabling truly real-time on-site genomic analysis for applications ranging from pathogen surveillance to microbial community profiling. 
\end{abstract}
\begin{document}

\flushbottom
\maketitle

\section*{Introduction}

\begin{figure}[!t]
    \centering
    \includegraphics[width=1 \linewidth]{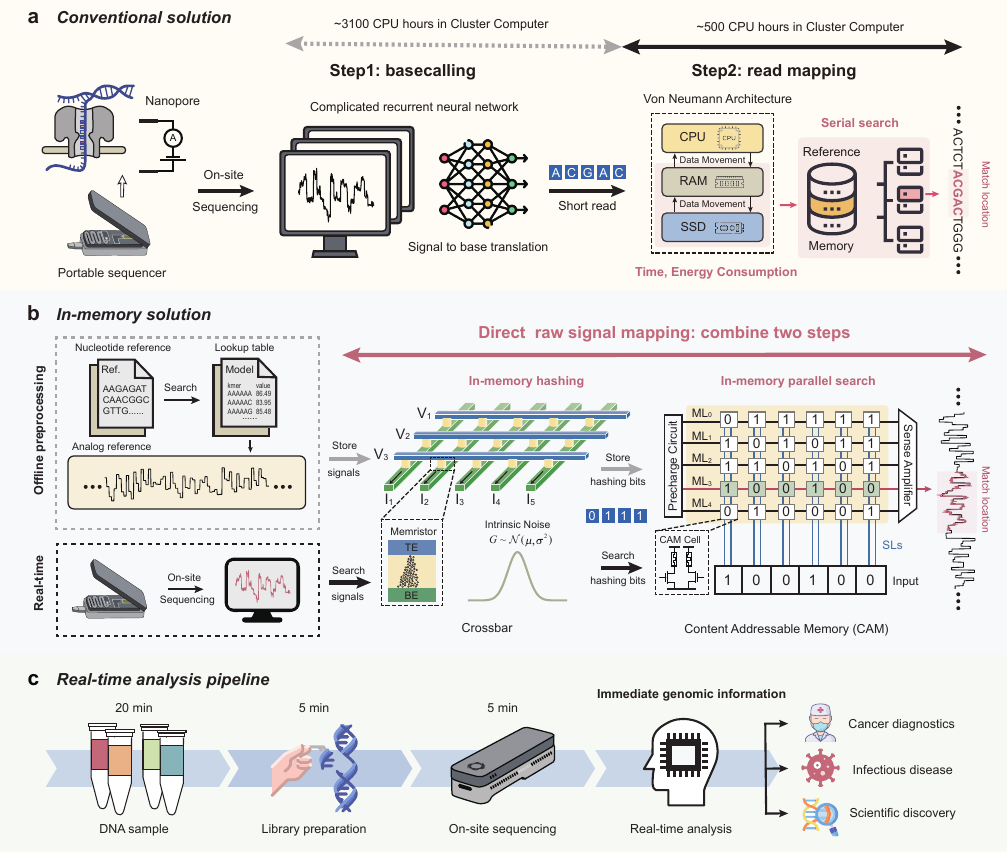}
    \caption{
        \textbf{Real time genomic analysis in memristor crossbar and memristor CAM.} 
        \textbf{a,} Conventional pipeline for nanopore genomic analysis. Nanopore sequencing generates noisy analog raw signals that must first be translated into precise nucleotide base pairs (basecalling). These short base pairs are then searched within a large reference genome database to identify their locations (read mapping). Due to the separate memory unit and compute unit of current von Neumann architecture, these two steps require frequent data movement, resulting in significant time and energy waste.
        \textbf{b,} Our in-memory computing software-hardware co-design for real-time genomic analysis. Our approach eliminates these inefficiencies by directly aligning analog raw signals with the analog reference genome. We hash the analog signals into binary features and perform parallel searches within memory, bypassing the need for separate basecalling and read mapping steps.
        \textbf{c} Efficient real-time analysis pipeline. This in-memory computing system delivers immediate genomic information with low latency and high energy efficiency, enabling rapid on-site sequencing and analysis for critical applications such as cancer diagnostics, infectious disease detection, and scientific discovery.
        TE, top electrode; BE, bottle electrode; ML, match line; SLs, search lines.
    }
    \label{fig1}
\end{figure}

Third-generation sequencing (TGS) technologies including those from Pacific Biosciences (PacBio) and Oxford Nanopore Technologies (ONT) have recently become highly popular sequencing platforms, which can generate a massive amount of ultra-long reads in parallel.
The high parallelism and longer reads have made these technologies widely used in \textit{de novo} assembly\cite{shafin2020nanopore}, RNA modifications detection\cite{garalde2018highly}, and other specialized applications\cite{wang2021nanopore}. 
Nanopore technologies, as commercialized by ONT, sequence the DNA and RNA by sensing the ionic current through the specific nanopore structure. Different molecules (a segment usually comprises 5-7 nucleotides) in the nanopore will lead to different current values, enabling the immediate analysis of raw current signals as they are generated. 
This real-time sequencing capability makes the nanopore sequencer a powerful tool for the detection of emerging infectious diseases\cite{virus1,virus2}, cancer diagnostics\cite{cancer}, and metagenomics\cite{metagenomic}.

Despite improvements in sequencing time, on-site, real-time genomic analysis has not yet been realized due to the extensive processing time and high energy consumption needed for high accuracy, significantly limiting their use in the scenarios described above. In detail, nanopore genomic analysis typically involves two computationally intensive steps: basecalling, which converts raw nanopore signals into nucleotides and read mapping, which determines the location of these translated reads in the reference genome (\figref{fig1}a). In a real-world nanopore sequencing study of human genomes\cite{bowden2019sequencing}, basecalling requires approximately 3,100 CPU hours, and read mapping takes around 500 CPU hours. 
A raw signal alignment method\cite{work1_loose2016real} has been proposed to bypass the complicated basecalling step (see details in Supplementary Fig. 2) and allow raw signals to provide richer analytical information than nucleotide sequences alone.
However, existing raw signal alignment methods\cite{work1_loose2016real, work2_uncalled, work3_readfish, work4_zhang2021real, work5_rawhash, work6_rawhash2, work7_rawalign, work8_sigmoni, work9_suigglenet, work10_squigglefilter,work11_shih2023efficient, work12_edwards2019real, work13_rawmap, work14_coriolis, work15_readbouncer, work16_deepselectnet} based on conventional hardware suffer from low parallelism and inefficiency as a result of separated memory unit and compute unit.
Specifically, fast and accurate alignment of analog raw signals with a large reference genome requires frequent and complicated similarity computations, causing significant time and energy overhead due to frequent data movement. 
Additionally, the low amplitude of the raw current signals, typically ranging from 60 to 120 picoamps (pA) for the R9.4 chemistry\cite{work2_uncalled}, makes them susceptible to noise-induced errors. This causes not only incorrect readings of current amplitude but also accidental addition and removal of signal segments, further complicating raw signal analysis.
These limitations are particularly problematic when deploying portable nanopore sequencers at the edge, which require minimal energy consumption and rapid processing for immediate genome information\cite{runtuwene2019site}, calling for hardware acceleration solutions.

In-memory computing based on emerging non-volatile memories provides higher storage density and energy efficiency, and has been widely explored across various data-intensive domains including artificial intelligence\cite{wang2019situ, yao2020fully, zhang2023edge, wan2022compute}, scientific computing\cite{le2018mixed, zidan2018general, song2024programming}, and optimization problems\cite{cai2020power, jiang2023efficient}. Thus, it should also be a strong candidate for real-time genomic analysis. Nevertheless, current in-memory solutions\cite{CIM_DNA1_zhang202465, CIM_DNA2_savi, CIM_DNA3_reconfigurable, CIM_DNA4_radar, CIM_DNA5_edam, CIM_DNA6_dash} utilizing emerging memories for sequence alignment focus solely on nucleotide-based read mapping, ignoring the intensive basecalling step\cite{shahroodi2023swordfish, simon2025cimba}, failing to meet practical requirements for real-time sequence analysis.
Furthermore, even when considering nucleotide alignment, the tolerance of high error rates in nanopore sequencing has not been experimentally validated in these simulation studies\cite{CIM_DNA2_savi, CIM_DNA3_reconfigurable, CIM_DNA4_radar, mao2022genpip} based on emerging memories with non-idealities, let alone the fact that raw signals are noisier and more complicated. 

This work introduces the first end-to-end experimental demonstration of memristor-based real-time genomic analysis, showcasing various applications such as infectious disease detection and metagenomic classification. 
Unlike conventional methods that determine the nucleotides sequence from raw signals (basecalling) and then compare them with the references (read mapping) in separate steps, we introduce a memristor-based heterogeneous in-memory computing hardware-software co-design that combines these two steps, enabling the immediate and efficient alignment of analog real-time raw signals as they are generated. 
On the software side, we propose a novel fuzzy seed-and-vote method for raw signal alignment, which employs a fuzzy seeding technique and relies on multiple search results to achieve high analysis accuracy. At the hardware level, the high parallelism of the memristor crossbar and content-addressable memory (CAM) is fully exploited to implement our new algorithm (\figref{fig1}b).
Leveraging from the intrinsic stochasticity of the memristor device, memristor crossbar arrays can efficiently perform locality-sensitive hashing (LSH) to achieve high dimensional feature extraction on noisy analog signals. Then the extracted feature is searched in parallel with CAMs,  which returns the corresponding location in one step, thereby avoiding the frequent data movement. To tolerate the large noise in raw signals, we custom designed the content addressable memory to perform approximate searches, allowing for variations in the feature vectors during the search process. 

Our approach was experimentally validated within a fully integrated memristor macro, showing 96.36\% F1 score in infectious disease detection and 98.88\% classification accuracy in metagenomic classification. We also analyzed the performance of our system based on experimentally-calibrated model in raw signal alignment task with real sequencing dataset, achieving software comparable 97.15\% F1 score with \SI{0.53}{\micro\second} searching latency and \SI{0.84}{\micro\joule} energy consumption per read. This presents a speed up of \textbf{$28,260\times$} compared to traditional CPU and \textbf{$51\times$} compared to a 28 nm ASIC, along with energy savings of \textbf{$370,864\times$} and \textbf{$477\times$} respectively.
Additional real-world applications including virus detection and relative abundance estimation were also simulated under real-world device variations, showing the robustness of this design. The successful demonstration of in-memory computing based raw signal alignment paves the way for an efficient real-time analysis pipeline, which integrates portable sequencers and neuromorphic chips to enable on-site sequencing and provide immediate genome information (\figref{fig1}c).

\section*{Results}
\subsection*{Hardware-software co-design for real-time genomic analysis}
\begin{figure}[!t]
    \centering
    \includegraphics[width=1 \linewidth]{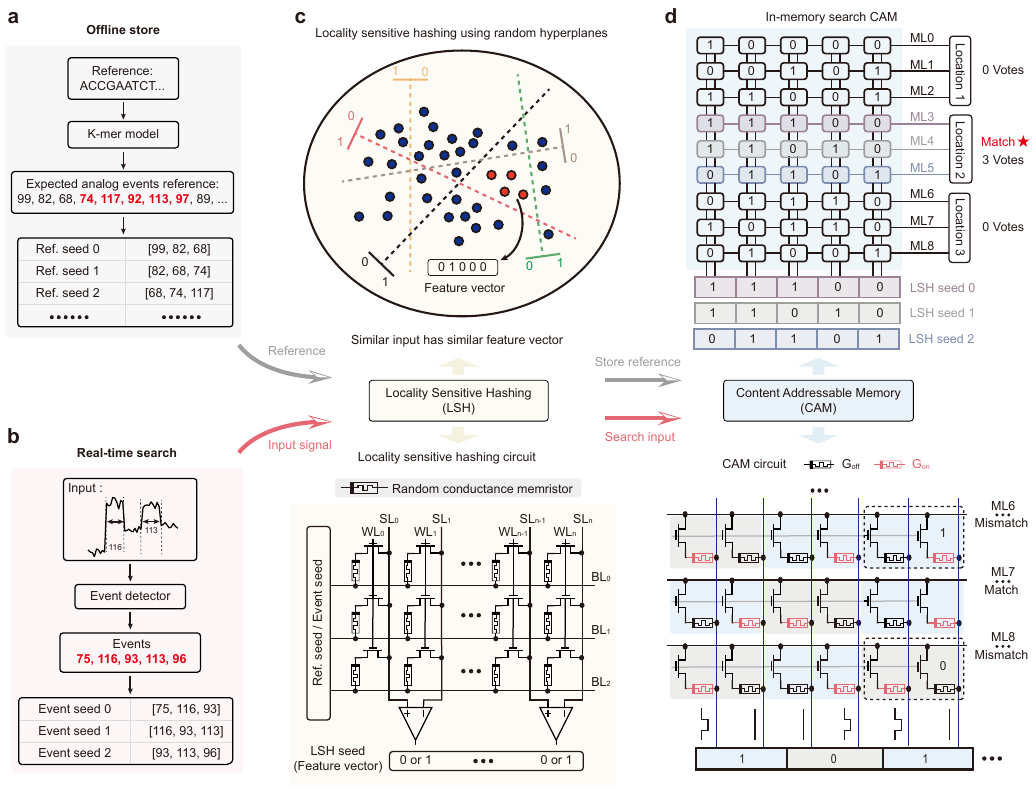}
    \caption{
        \textbf{In-memory raw signal analysis pipeline with hardware-software co-design.} 
        \textbf{a,} Reference data preprocessing. According to the \textit{k}-mer model released by ONT, the reference is converted into expected analog vectors representing the ideal currents when the given reference DNA sequences are in the nanopore.
        \textbf{b,} Input signal data preprocessing. The analog event value is calculated from the mean of regions between two sudden change boundaries detected by t-test. Each boundary approximately indicates that a new DNA molecule goes through the nanopore. This rough and simple segmentation will introduce many errors, which should be tolerated in following steps.
        \textbf{c,} Locality sensitive hashing via random hyperplanes and its hardware implementation. Analog vectors are transformed into binary feature vectors by projecting them onto many random hyperplanes, with their positions relative to each hyperplane determining the binary output. Random memristor crossbar arrays can efficiently implement this algorithm by utilizing the intrinsic noise of memristors. 
        \textbf{d,} Approximate in-memory search engine, content addressable memory, performs fuzzy seed-and-vote algorithm. The extracted feature vectors from the reference genome are sequentially stored in each memory row, with two different conductance states in memristor encoding a binary `0' or `1'. After comparing all input LSH seeds with all the stored memory, the matched location should accumulate obviously more votes than other locations.
    }
    \label{fig2}
\end{figure}

\figref{fig2} illustrates our hardware-software co-designed implementation for real-time genomic analysis. 
Instead of following the traditional two-step approach of basecalling (converting the sequencer's raw current signals to nucleotide sequences) followed by read mapping (aligning sequences to the reference genome), we directly align the raw signals by searching the locality hashed bits using low-cost hardware against those from reference genome in memory.
This approach offers two key advantages: the hashing bit is a direct representation of the raw signal and offer better compressibility and noise / error tolerance compared to nucleotide sequences for real-time analysis, and the in-memory hashing and search eliminate frequent data movement between memory and compute units, potentially delivering significant improvements in speed and power efficiency through hardware-software synergy.

The implementation of this raw signal analysis pipeline, however, presents several challenges.
First, searching against massive amounts of data is extremely computationally expensive, especially given the significant noise and errors in the raw signals.
The conventional genomic analysis pipeline's alignment step is already time- and energy-intensive because basecalled nucleotide sequences contain multiple error types. Insertion and deletion errors are more challenging to handle than substitution errors and typically require dynamic programming algorithms, which are commonly more computational intensive than the other string matching algorithms and difficult to accelerate because of low parallelism.
Our direct raw signal search is even more challenging as these signals are inherently noisier and more complex analog signals, with more sophisticated error patterns\cite{work2_uncalled} that vary based on preprocessing methods. 
Therefore, the preprocessing stage must be carefully optimized to prepare raw signals specifically for the in-memory search capabilities of our hardware.
Second, given the signal is far noisier than the basecalled nucleotide sequences, it is unlikely to find exact matches in the memory. 
This necessitates custom-designed in-memory searching hardware capable of approximate search, with parameters specifically tuned to complement our preprocessing approach. The interdependence between hardware capabilities and software preprocessing demonstrates why this work demands meticulous hardware-software co-design to create a viable pipeline.

\subsubsection*{Optimize raw signal preprocessing for in-memory search}
For nanopore sequencing technology, the raw signal is the current measured as genomic DNA or RNA sequences pass through the nanopore\cite{david2017nanocall}.
This current is influenced by both the nucleotide directly in the pore and its surrounding nucleotides, with diminishing effects from more distant bases.
To address this context dependence, researchers have developed a \textit{k}-mer model that predicts expected current values for specific \textit{k}-mer DNA or RNA sequences in the nanopore.
At its core, this model functions as a lookup table—for instance, when nucleotides `ACCGAA' occupy the nanopore, the expected current should be \SI{99.07}{\pico\ampere} (Fig. \ref{fig2}a).
When the sequencer detects significant changes in current, it indicates that a new nucleotide has moved into the pore, creating what we define as an event.
The basecaller then consults this lookup table to determine the most probable sequence corresponding to each current reading—if \SI{99.07}{\pico\ampere} is detected, the nucleotides `ACCGAA' are likely present in the pore (Fig. \ref{fig2}b).
This mapping, however, is rarely one-to-one, as different \textit{k}-mers can produce nearly identical current values.
Consequently, the basecaller employs sophisticated models that integrate readings from multiple sequential events to reconstruct the most probable nucleotide sequence.
To maximize sensitivity, these algorithms typically use low thresholds for event detection, which effectively capture true transitions but also misidentify noise as events. This approach inherently generates more `stay' errors (appearing as insertions in the output sequence) than `skip' errors (appearing as deletions).
While this bias is acceptable for standard sequencing applications, where exact nucleotide sequences need to be extracted, it creates significant challenges for comparative analyses against reference sequences, particularly in edge computing and field-based genomic applications where exact sequence determination is less critical than pattern matching.

To optimize for direct comparison of raw signals against the reference, we enhance the event detection to recognize only representative events where current reads differ noticeably from adjacent events.
The event detector is modified from several previous studies\cite{work2_uncalled, work4_zhang2021real, work5_rawhash, work6_rawhash2}, which use t-tests within sliding windows to detect sudden signal changes.
After this step, the read sequence is converted to a high-dimensional vector with each dimension representing an average current read during an event corresponding to a \textit{k}-mer DNA molecule.
Through our modifications (more details can be found in Methods), we achieve a relatively balanced number of `stay' and `skip' errors, which is crucial for hardware implementation in subsequent steps.
The reference sequences for comparison are also converted to expected current read vectors using the \textit{k}-mer model.
The raw signal alignment then focuses on locating the read vector within the reference vector. 

\subsubsection*{Fuzzy seed-and-vote in memristors}
We adopt the seed-and-vote algorithm\cite{liao2013subread, liu2016fast} to locate the read vector within the reference vector, which follows the principle that if read genomic events are adjacent, they should also be adjacent in the reference genome.
In this algorithm, the read vector is broken into short overlapping fragments (seeds), and these seeds are independently mapped to potential reference vector locations. 
A voting mechanism is then used to determine the most likely location of the read vector within the reference vector.
The original algorithm mapped digital genomic nucleotide sequences, but this creates a challenge when matching raw signals (mean current values during an event) because exact matches don't exist here.
Therefore, we introduce a fuzzy seed-and-vote algorithm to tolerate noise and errors in the raw signals, where if the read vector is close enough to the reference vector, it's still considered a match.
Searching analog vectors in such a massive space is extremely computationally expensive, so the seeds (both read and reference) are hashed into binary feature vectors using memristor-based locality sensitive hashing (LSH) to reduce the search space (Fig. \ref{fig2}c).
The LSH seeds are then searched in the memristor content addressable memory (CAM) to find the most likely location of the read vector within the reference vector (Fig. \ref{fig2}d).

The LSH is implemented in memristor crossbar arrays, exploiting intrinsic stochasticity in the programming of devices, along with non-volatile retention after the programming. This not only reduces the search space but also generates binary hashing bits that can be efficiently searched in the CAM hardware. 
The idea is that after hashing, the binary hashing bits maintain the same distance property as the original analog vectors, so similar analog event vectors will be hashed into similar binary feature vectors.
To implement LSH, the input analog event vectors are projected onto many random hyperplanes, and the binary output is determined by their positions relative to these hyperplanes. 
This hashing method is mathematically implemented by the following equation: $\mathbf{h} = H(\mathbf{a} \cdot \mathbf{G})$, where $\mathbf{a}$ denotes the input analog event vector seeds, $\mathbf{G}$ is a random matrix with normally distributed elements, $H(\cdot)$ is the Heaviside step function, and $\mathbf{h}$ represents the hashed bits.
The matrix multiplication of the input seed and random matrix can be efficiently implemented in memristor crossbar arrays, leveraging the intrinsic stochasticity of memristor devices.
Since memristor conductance in the crossbar array follows the same mean value after identical writing processes (e.g., after resetting with the same parameter or iterative write-and-verify with the same target conductance), the random matrix $\mathbf{G}$ can be implemented by subtracting two adjacent columns, naturally creating a zero-mean distribution.
The Heaviside step function is implemented using a comparator connecting adjacent columns; it outputs `1' when the odd column's sum current exceeds the even column's, and `0' otherwise. Unlike traditional matrix multiplication where memristor stochasticity must be minimized through write-and-verify programming, this approach utilizes the intrinsic noise of memristors to facilitate locality sensitive hashing, avoiding time-consuming programming by using simple reset pulses.
After the hashing process, the generated hashing bits represent the binary feature of the seed while preserving the distance properties of the original analog event seeds, enabling efficient search in the CAM hardware.

After the hashing bits of a seed during a read are generated, they are compared to the hashing bits of the seeds in reference that are stored in a memristor CAM.
CAM is a memory hardware that takes a data pattern as input and returns the address of the memory cells that store the pattern, which has been widely used in many data-intensive applications, such as artificial intelligence\cite{ni2019ferroelectric} and genomic analysis\cite{CIM_DNA2_savi}.
However, previous CAM implementations are usually designed for exact matching, which are not suitable for our application due to noise and errors in the raw signals.
Therefore, we co-design the CAM to perform approximate searches, where the CAM will consider a seed a match even if there are several bits mismatches.
The design of the memristor-CAM is shown in Fig. \ref{fig2}d, where each memristor-CAM cell consists of two memristors that store a binary bit with a differential encoding. 
When the search operation begins, the input hashed bits are applied to the search lines as a differential voltage. 
If the search data matches with the stored data, only the memristor in the high resistance state will be activated, contributing a negligible current; while if there is a mismatch, the memristor in the low resistance state will be activated, contributing a detectable current value. 
The current value will be summed up along the match line according to Kirchhoff's current law, and therefore the resulting current represents the number of bit mismatches between the input and the stored data. 
A sense amplifier with a tuneable threshold is used to output the match result, where the threshold is set to tolerate a certain number of mismatches.

If a search for a seed from a read returns a match in the CAM, indicating the mismatch bits are below the threshold, the corresponding location accumulates one vote.
After searching all seeds in a read, the location with significantly more votes than other places, having more than twice the votes of any other location, is identified as the position of the read in the reference genome.

This approach doesn't rely on a single search result but aggregates results from multiple search outcomes, enhancing robustness against non-ideal factors in sequencing machines and memristors.
It ensures that even if individual searches are affected by noise or defects, the cumulative voting process can still identify the most probable position within the reference sequence, improving accuracy and reliability.
A detailed example can be found in Supplementary Fig. 3.
To sum up, the successful mapping of noisy raw signals can be attributed to the fuzzy seeding technique and multiple searching (see more details in Methods).
\subsection*{Raw signal genomic analysis experiments on integrated memristor hardware}
\begin{figure}[!t]
    \centering
    \includegraphics[width=\linewidth]{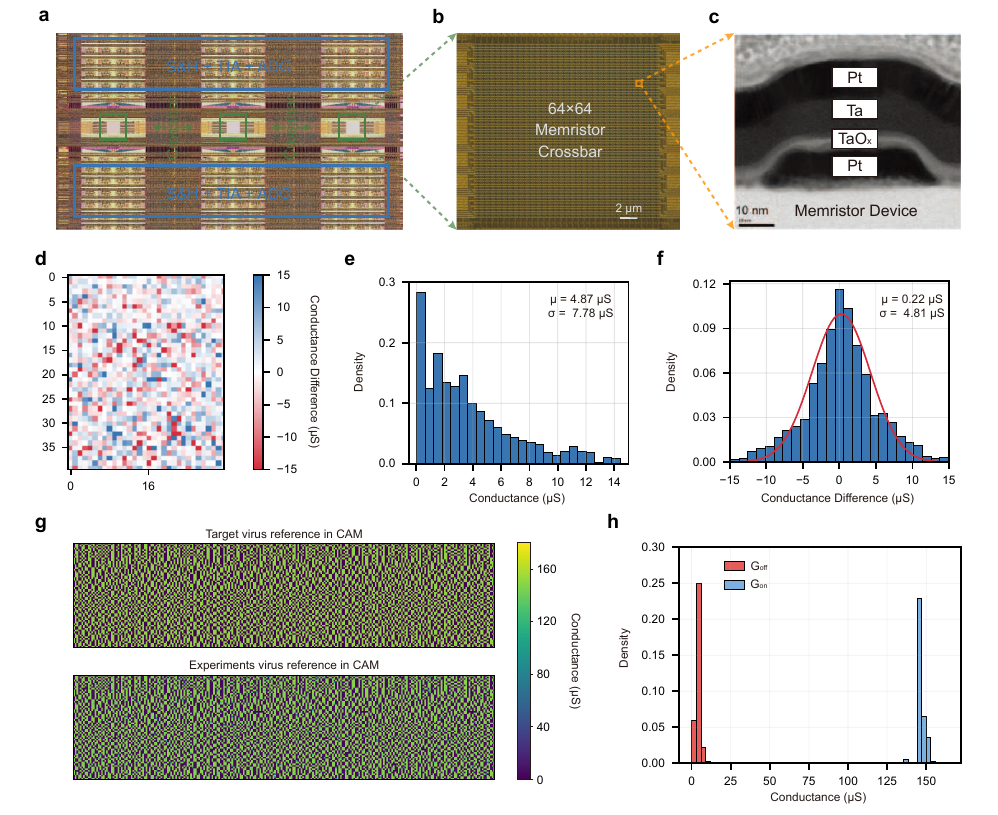}
    \caption{
        \textbf{Experimental implementation for memristor based real-time genomic analysis.} 
        \textbf{a,} The figure of our fully integrated memristor chip with digital and analog peripheral circuits.
        \textbf{b,} The optical figure of one memristor crossbar array.
        \textbf{c,} The cross section TEM figure of one nanoscale memristor device.
        \textbf{d,} The zero-mean $40 \times 32$ random array generated by subtracting adjacent columns of a $40 \times 64$ conductance matrix.
        \textbf{e,} The random conductance distribution of our memristor arrays.
        \textbf{f,} The corresponding random conductance difference distribution in our memristor arrays with rough zero-mean conductance.
        \textbf{g,} The ideal target virus conductance $64 \times 256$ array of CAM and experimental $64 \times 256$ virus conductance array of CAM.
        \textbf{h,} The detailed conductance distribution of binary experimental virus CAM array.
    }
    \label{fig3}
\end{figure}

The proposed hardware-software co-designed approach has been experimentally demonstrated in a 180 nm fully integrated memristor crossbar chip,  containing three $64 \times 64$ crossbar arrays and peripheral circuits including S\&H, TIA, and ADC, among other modules (\figref{fig3}a). The optical image of one crossbar array has been shown in  \figref{fig3}b, where the 50 nm $\times$ 50 nm memristors are integrated using back-end-of-the-line (BEOL) processing on the top of CMOS circuits. \figref{fig3}c presents a cross-sectional TEM image of an integrated Ta/\ce{TaO_x}/Pt memristor, with \ce{TaO_x} functioning as the resistive switching layer.

The integrated memristor platform demonstrates the capability to implement the LSH algorithm and approximate matching CAM operations. 
For LSH, as shown in \figref{fig2}a-d, we need to generate a zero-mean random matrix, which can be implemented through write operations on the memristor crossbars. 
By applying several identical reset pulses to all devices in the array, we generate a $40 \times 64$ random conductance matrix, with experimental readout results shown in \figref{fig3}d and \figref{fig3}e.
The algorithm's required zero-mean distribution is achieved by our design where each matrix element is represented by the conductance difference between two memristor devices, as confirmed in \figref{fig3}f.
The randomness is naturally programmed by exploiting the inherent variability of memristor devices, eliminating the need for additional expensive random number generators.
The platform also successfully performs approximate matching CAM operations by programming LSH-hashed seeds from virus reference genomic sequences into the memristor CAM arrays. 
\figref{fig3}g presents both the target conductance values representing reference sequences and the experimental readout values after programming to our memristor CAMs, while \figref{fig3}h shows their distribution.
The programmed CAM array exhibits a binary-like conductance distribution with minimal variation, demonstrating close alignment with target values. 
Additional programming results for other species references are available in Supplementary Fig. 4.

\subsubsection*{Infectious disease detection experiment}

\begin{figure}[!t]
    \centering
    \includegraphics[width=1 \linewidth]{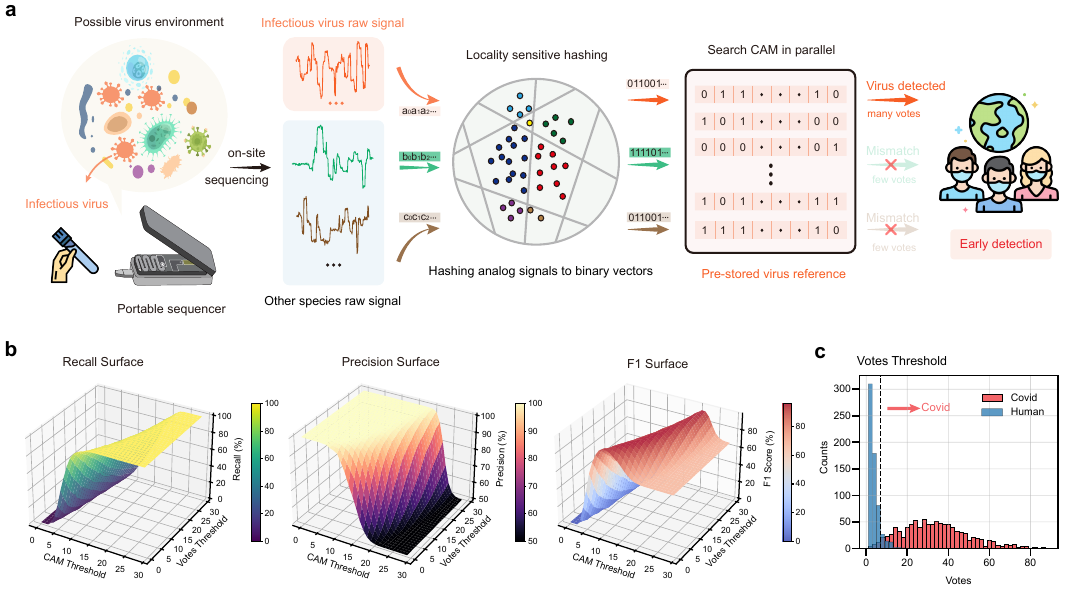}
    \caption{
        \textbf{Experimental results for infectious virus detection.}
        \textbf{a,} Infectious disease detection pipeline. Scientists sequence all the species in the potential virus environment and align all raw signals to pre-stored virus reference on-site for real-time analysis. Once emerging infectious virus is detected, strategies should be promptly implemented to prevent economic and human losses.
        \textbf{b,} The recall, precision and F1 surface as CAM threshold and votes threshold. The best F1 score 96.36\% is achieved when the recall and precision are well balanced.
        \textbf{c,} The distribution of votes with a fixed CAM threshold of 16. Reads that match the pre-stored virus reference will have more votes, so reads with votes larger than seven will be regarded as COVID virus.
    }
    \label{fig4}
\end{figure}

We first showcase the capability of our memristor-based solution for infectious disease detection.
The outbreak of COVID-19 in 2020 has caused around 7 million deaths worldwide and led to an economic loss exceeding US\$22 trillion by 2025\cite{das2024systems}.  
Developing early detection systems for monitoring emerging infectious diseases is therefore critical.
Current detection methods rely heavily on Polymerase Chain Reaction (PCR), which has several limitations: developing disease-specific primers takes months, the process requires bulky equipment, and new variants like Omicron may evade detection.
While sequencing technology has advanced with miniaturized devices, the enormous computational power required for sequence mapping has prevented the widespread adoption of more sensitive and accurate sequencing-based pathogen detection.
Our memristor-based system addresses this computational bottleneck, enabling a real-time infectious disease detection pipeline that can be deployed on-site for rapid response (\figref{fig4}a).
Scientists collect samples from high-risk viral environments and complete sequencing with portable sequencing machines. 
Our system then efficiently and accurately maps the raw nanopore current signals to virus reference genomes stored in our system, determining if infectious diseases are present in real-time.
This immediate identification allows appropriate containment strategies to be implemented during the earliest stages of a potential pandemic, significantly reducing both mortality and economic impact.

Considering the limited memristor count in our prototype system, we selected a random 78 base pair (bp) fragment of SARS-CoV-2 as our virus reference genome for proof-of-concept, while more extensive simulations based on the experiments with real sequencing data are discussed in subsequent sections.
This 78 bp fragment generates 73 expected analog current reads, each corresponding to an event when nucleotides pass through the nanopore according to the \textit{k}-mer model ($k$ = 6 in this experiment). 
To process this data, current reads from neighboring 10 events are grouped together to form one reference seed, which is then hashed into a 128-bit binary vector using LSH implemented on four 10$\times$64 memristor crossbar arrays with random conductance values.
Therefore, the 73 expected analog current reads are hashed into 128 binary vectors and stored in the $64 \times 256$ array in our system, as experimentally demonstrated in \figref{fig3}g.
To simulate real-world virus detection conditions, we created an artificial community containing 1,000 simulated reads in raw signals format from the same 78 bp virus reference genome and another 1,000 reads corresponding to random 78 bp from human chromosome 1, using the raw signal simulator Squigulator\cite{gamaarachchi2024simulation}.
During parallel search operations, our system identifies approximate matches when the hamming distance between a hashed input and reference bits stored in any CAM row falls below a predefined threshold, adding one vote for that location. 
After processing all input vectors, if the accumulated vote count exceeds a specified threshold, the system determines that the input raw signal successfully aligns with the stored virus reference (\figref{fig4}a), indicating virus detection.

The classification results comparing the raw signal reads of virus and human genomic sequences are shown in \figref{fig4}b. 
We define true positives (TP) as virus reference genome signals correctly classified as virus, false negatives (FN) as virus signals misclassified as non-virus, and false positives (FP) as human reference genome signals incorrectly classified as virus. 
To evaluate classification performance, we calculate:
\begin{equation}
    \text{Recall} = \frac{\text{TP}}{\text{TP} + \text{FN}}, \quad \text{Precision} = \frac{\text{TP}}{\text{TP} + \text{FP}}, \quad \text{F1} = \frac{2 \times \text{Recall} \times \text{Precision}}{\text{Recall} + \text{Precision}}
\end{equation}
Our analysis reveals a trade-off: as the approximate match CAM threshold increases, recall improves but precision decreases due to rising false positives. 
The optimal balance occurs at a CAM threshold of 16 and votes threshold of seven, yielding a peak F1 score of 96.36\%. 
\figref{fig4}c shows the vote distribution with this fixed CAM threshold of 16, where signals receiving more than seven votes are classified as SARS-CoV-2 virus. 
The clear separation between virus and non-virus vote counts explains the high classification accuracy achieved by our system.
Applications prioritizing sensitivity should use higher CAM thresholds with lower votes thresholds, while those requiring specificity should employ lower CAM thresholds with higher votes thresholds.

\subsubsection*{Metagenomic classification experiment}

\begin{figure}[!t]
    \centering
    \includegraphics[width=1 \linewidth]{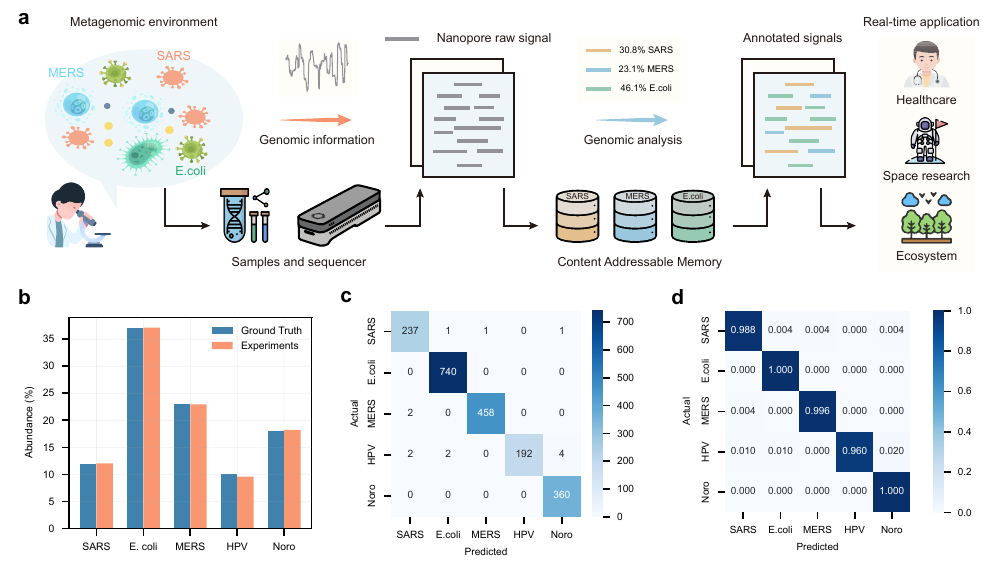}
    \caption{
        \textbf{Experimental results for metagenomic classification.}
        \textbf{a,} On-site relative abundance estimation. Raw signals in metagenomics environment will be efficiently aligned to pre-stored genome references. Each raw signal will be accurately assigned to its corresponding species. 
        \textbf{b,} The true abundance and experimentally estimated abundance in relative abundance estimation experiments. 
        \textbf{c,} The detailed confusion matrix of relative abundance estimation experiments.
        \textbf{d,} The normalized confusion matrix of relative abundance estimation experiments.
    }
    \label{fig5}
\end{figure}

In addition to detecting viral presence, our memristor-based raw-signal genomic analysis system can also estimate the relative abundance of different species in a metagenomic environment.
Real-time genomic analysis using portable sequencing machines is crucial for on-site metagenomics, particularly where samples are difficult to preserve and immediate genomic information is needed\cite{runtuwene2019site}. \figref{fig5}a illustrates the implementation of an in-memory computing based on-site metagenomic analysis pipeline. The mapping results reveal the microbial composition of the environment, supporting disease diagnosis\cite{liu2025analysis}, ecosystem monitoring\cite{liu2025analysis} and space exploration\cite{wang2021nanopore}.

The hardware implementation for estimating relative abundance is similar to that used in infectious disease detection. The key difference is that virus detection focuses on determining the presence of an emerging virus, whereas here we quantify the proportion of each species in the environment. To be more specific, we randomly selected five 78 bp fragments from five different kinds of microorganisms as reference genomes, and created an artificial microorganisms community that includes 2,000 artificial reads in raw signals format generated from these reference genomes utilizing the Squigulator\cite{gamaarachchi2024simulation}. The five virus reference CAM arrays were also generated in the same way as for virus detection. Once parallel search is complete, the input raw signals will be assigned to the reference species that receives maximum votes.
As shown in \figref{fig5}b, blue bars represent ground truth abundance while orange bars show our experimentally estimated abundance. 
We obtained these results using a CAM threshold of 16, which corresponds to the threshold yielding the highest F1 scores in virus detection. 
The detailed original and normalized confusion matrices (\figref{fig5}c,d) demonstrate that our system achieves 98.88\% accuracy in correctly classifying raw signals.

Metagenomic classification presents greater scalability challenges than virus detection because it requires pre-storing genomes of all environmental species rather than just a single reference genome. 
One way to address this challenge is by implementing the minimizer technique\cite{roberts2004reducing, work6_rawhash2}, which selects representative \textit{k}-mers as input and reference instead of using all \textit{k}-mers. 
This approach can significantly reduce the number of reference \textit{k}-mers that need to be stored in CAM, saving both time and energy during searches. 
While this data reduction might decrease performance due to noise in raw signal data, we can mitigate this effect by incorporating more advanced neural network based event detectors in the future.
\subsection*{Performance benchmark on scaled architecture}

\begin{figure}[!t]
    \centering
    \includegraphics[width=1 \linewidth]{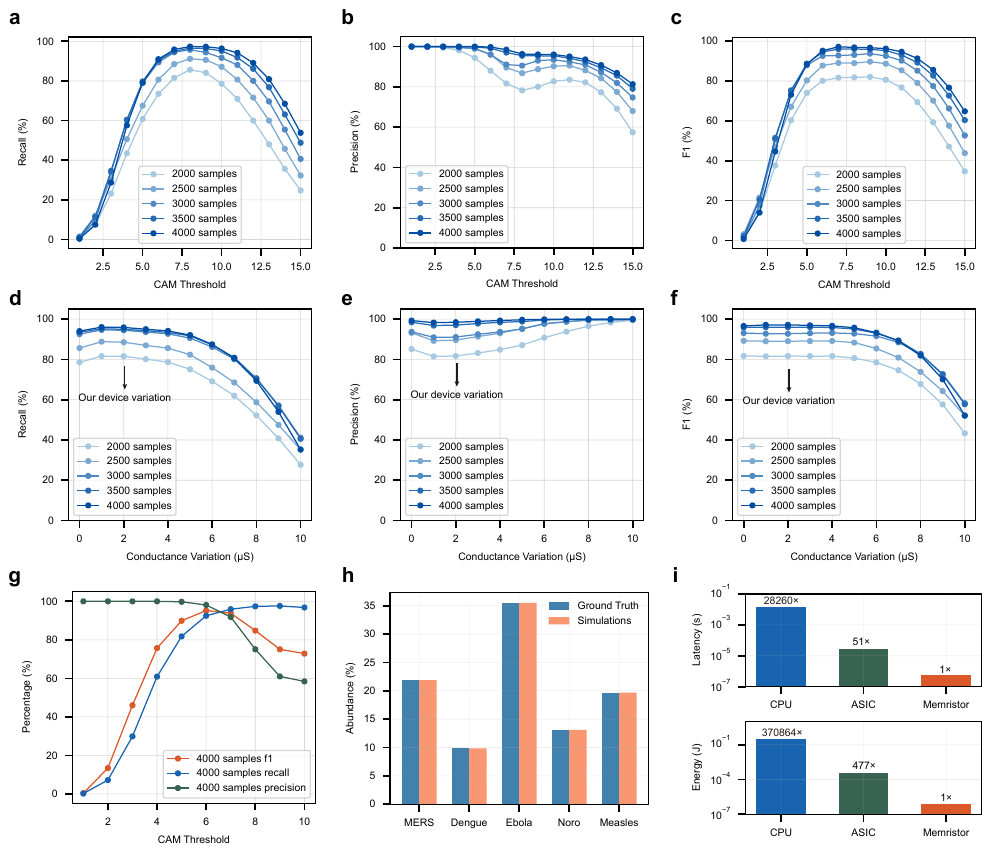}
    \caption{
        \textbf{Scaled-up simulations and performance comparison.} 
        \textbf{a-c,} Recall, precision, and F1 scores as a function of CAM threshold and sampling number in the SARS-CoV-2 read mapping task based on the experimental result.
        \textbf{d-f,} Recall, precision, and F1 scores as a function of conductance variation when CAM threshold is fixed at seven, and the experimental conductance variation is shown with arrows.
        \textbf{g,} Scaled-up virus detection performance simulations based on the experimental result. 
        \textbf{h,} Scaled-up relative abundance estimation simulations based on the experimental result.
        \textbf{i,} Latency and energy per read comparison with CPU and ASIC based on the Von Neumann architecture.
    }
    \label{fig6}
\end{figure}

Although these two experiments demonstrate that our hardware-software co-designed memristor-based hardware can efficiently and effectively analyze genomic raw sequence signals, limitations remain. 
First, due to the limitation of our prototype system, the reference genome for each species in our experiments is very small with only one location. 
Additionally, the raw signals are generated by Squigulator which may not fully represent the real-world nanopore sequencing data.
To better evaluate our memristor-based software-hardware co-design for raw genomic signal analysis, we conduct simulations based on our experiment-calibrated model using complete reference genome and real nanopore R9.4.1 sequencing data. 
The locality sensitive hashing conductance is obtained from the memristor chip, and the conductance values for approximate-match CAM were sampled from the experimental conductance distribution, ignoring obvious outliers.

Here, we first analyze the read mapping task to benchmark the performance of our memristor-based raw signal alignment, as this is a common benchmark in algorithm studies. 
Our experimentally-validated simulation tool enables mapping of the entire SARS-CoV-2 reference genome in real nanopore R9.4.1 raw signals format, whereas earlier experiments could only store small portions of the reference genome due to hardware limitations. 
In the simulation, we map 10,000 reads from real R9.4.1 SARS-CoV-2 raw signals, each representing a fragment of the SARS-CoV-2 genome, to the SARS-CoV-2 reference genome.
If the location with the maximum votes has at least twice as many votes as the second-highest location, and the maximum votes exceed a certain minimum threshold, we consider the raw signal to belong to the reference genome of this location.
The recall, precision, and F1 score for mapping to the correct location are shown in \figref{fig6}a-c.
As the CAM threshold increases, recall initially rises and then falls; precision first decreases, then increases, and subsequently decreases again. 
This non-monotonic behavior can be explained by competing effects of the CAM threshold. 
At very low thresholds, correct inputs rarely match with corresponding reference, resulting in low recall; 
As the threshold increases, more correct matches are identified, improving recall. 
However, when the threshold becomes too large, the specificity of matching decreases, causing inputs to match with incorrect locations. 
This reduces the vote differential between correct and incorrect locations, leading to more false negatives and fewer true positives.
The optimal balance between recall and precision occurs at a CAM threshold of seven, yielding an F1 score of 97.15\% with 4,000 raw signal samples per read, taking into consideration device variation of our analog hardware.
This F1 score is comparable to the state-of-the-art raw signal alignment solution implemented in digital hardware, UNCALLED\cite{work2_uncalled}, which has an F1 score of 97.34\% in the same task.

Our design not only offers tolerance to noise and errors in the sequencing data but is robust to device non-idealities in the memristor crossbar, which is key for successful hardware implementations.
\figref{fig6}d-f show the memristor device variation study, where the performance is compared with varying conductance variation when the CAM threshold is seven.
The results demonstrate that the F1 score remains stable even when the standard deviation of the device variation reaches to \SI{5}{\micro\siemens}, which is more than twice our experimental results.
The robustness of this design can be attributed to two key factors: 
first, our method aggregates multiple search results rather than relying on a single match, providing resilience against individual matching errors; 
second, we adopt an approximate match approach during the search, so that small variations do not impact the final result, as evidenced by the consistently high F1 scores across CAM thresholds between 7 and 10.

Another key advantage of raw signal analysis is the ability to stop sequencing when we detect that a read is meaningless or when sufficient information has been collected to make a decision, saving both time and costs.
This capability is commercially available in nanopore sequencers as the "Read Until" functionality.
To optimize this feature, we investigated how the number of raw-signal current samples affects our alignment performance.
Nanopore sequencers with R9.4.1 chemistry operate at a sampling frequency of 4,000 Hz, processing approximately 450 base pairs per second.
Our results show that performance metrics—recall, precision, and F1 score—all improve with increasing sample numbers, as illustrated in \figref{fig6}a-c. 
These metrics plateau at approximately 4,000 samples per read, equivalent to 1 second of sequencing time. 
This finding indicates that 4,000 samples represent an optimal balance between accuracy and speed for raw signal alignment, allowing early termination of sequencing when appropriate while maintaining reliable results.

Finally, we conduct scaled-up simulations for real-world applications (\figref{fig6}), for virus detection and relative abundance estimation, respectively, using the same method as in previous end-to-end experiments.
For virus detection, we randomly selected 1,000 R9.4.1 virus raw signals and 2,000 R9.4.1 E. coli raw signals from a real sequencing dataset as input, with the virus reference stored in CAM. 
When setting the CAM threshold to 6, we achieved the best F1 score of 95.21\%.
We specifically chose a lower threshold because precision is more critical than recall for virus detection applications compared to read mapping task introduced in the previous paragraph.
For relative abundance estimation, 1,000 raw signals from different species are generated by Squigulator\cite{gamaarachchi2024simulation} with default settings from random positions in the reference genome, and the ground truth abundance and estimated abundance are also shown in \figref{fig6}h with 99.77\% F1 scores.
These results demonstrate high accuracy of our approach for real-world genomic analysis tasks.

\figref{fig6}h compares the latency and energy consumption of our memristor-based genomic analysis in the SARS-CoV-2 raw signal reading mapping task with UNCALLED\cite{work2_uncalled} using FM-index\cite{simpson2010efficient} on CPU, and SquiggleFilter\cite{work10_squigglefilter} using dynamic time warping\cite{berndt1994using} on ASIC. 
Our simulation considers the overhead of the peripheral circuit, such as DAC, TIA and other peripheral circuits for memristor arrays.
Details about how to calculate numbers in the comparison can be found in Supplementary Note 1. 
Our result shows that the latency per read of our proposed memristor based method is \SI{0.53}{\micro\second}, which is \textbf{$28,260\times$} faster than traditional CPU and \textbf{$51\times$} faster than ASIC based on the Von Neumann architecture. The energy consumption for each read is \SI{0.84}{\micro\joule}, achieving \textbf{$370,864\times$} and \textbf{$477\times$} energy saving respectively. Our method's analysis speed significantly surpasses sequencing speed, enabling real-time analysis of up to 1.8 million nanopores simultaneously. This is 878 times more than the current MinION technology, which supports only 2,048 nanopores \cite{wang2021nanopore} and leave space for future nanopore throughput increase.
This improvement comes from the synergy of software and hardware innovations. Software-wise, we directly analyze the raw sequencer signal rather than conducting two separated basecalling and read mapping, bypassing the unnecessary nucleotide representation. 
The dimension reduction using LSH and approximate matching steps and their hyperparameters are tailored for our hardware. 
Hardware-wise, the hashing operation fully exploits the inherent stochasticity of our emerging device, while approximately matching tolerant the device error. 
Both steps perform massively parallel analog operations directly in the memory device, removing expensive data transfer overhead.

\section*{Discussion}
In this work, we propose a memristor-based in-memory computing hardware-software co-design for fast, accurate, and efficient real-time raw signal genomic analysis at the edge. A fuzzy seed-and-vote algorithm is proposed to exploit the effective locality sensitive hashing capability in memristor crossbar and parallel search capability in memristive content addressable memory, combining the basecalling and read mapping steps. This method has been experimentally validated in infectious disease detection and relative abundance estimation tasks.
In extended simulations, our system achieved a \textbf{$51\times$} speed increase and a \textbf{$477\times$} improvement in energy efficiency compared to 28 nm ASICs, while being capable of simultaneously analyzing signals from up to 1.8 million nanopores in real-time.
Besides, the conductance variation analysis is also conducted, and its strong robustness to variation shows great potential for practical applications. This work breaks through the memory wall constraints, laying a solid foundation for the next generation of edge sequencing hardware and algorithms, and also opens the possibilities for other memory bottleneck bioinformatics algorithms, which also need the frequent comparison like metagenomics, RNA quantification, and protein alignment.

\section*{Methods}

\subsection*{Memristor integration with CMOS circuits}
The CMOS-compatible $\mathrm{Ta/TaO_x/Pt}$ memristors in our chip are monolithically integrated on top of CMOS chips with a standard 180 nm technology node in a back-end-of-the-line (BEOL) process. The fabrication process begins with the removal of the chip's passivation layer using reactive ion etching (RIE) and a buffered oxide etch (BOE) dip. Then a bottom electrode is formed by sputtering and patterning 2 nm of Cr and 10 nm of Pt using e-beam lithography.  A 2 nm layer of $\mathrm{TaO_x}$ is deposited via reactive sputtering as the resistive switching layer, which is followed by sputtering 10 nm of Ta as the top electrode, with an additional Pt layer for passivation and enhanced electrical conduction.

\subsection*{CAM programming with controlled stochasticity}
Two binary memristors are used to encode one bit in the CAM array, thus we need to program the on state conductance and off state conductance accurately in our crossbar array according to the reference genome. To minimize the static current consumption, we choose \SI{0}{\micro\siemens} and \SI{150}{\micro\siemens} as the target conductance of the $\mathrm{G_{off}}$ and $\mathrm{G_{on}}$. An iterative write-and-verify algorithm is utilized to program the conductance of memristors. We first set a tolerant programming error of \SI{5}{\micro\siemens} and iteratively tune the conductance of the entire array to their target conductance values by applying SET and RESET pulses with a width of \SI{1}{\micro\second}. In each iteration, we first use a 0.2 V READ pulse to measure the conductance. If the current conductance exceeds the pre-defined tolerant error of \SI{5}{\micro\siemens} above the target, a reset pulse with sequentially increasing amplitude will be applied to the memristor. Conversely, if the conductance is more than \SI{5}{\micro\siemens} below the target, a set pulse will be applied. If the conductance is within the tolerant range, the programming algorithm for the current memristor will be terminated.

\subsection*{LSH programming utilizing intrinsic noise}
The intrinsic stochastic switching behavior of memristor devices leads to a roughly lognormal conductance distribution in memristor arrays. Unlike CAM programming, where the stochasticity of memristors needs to be controlled, we utilize this intrinsic noise to generate random arrays for efficient hashing. We reset the whole crossbar array with five 1.5 V RESET pulses, each with a width of 20 ns. Due to some non-ideal influences, some conductance values may become extremely high. Therefore, we program the entire memristor arrays to \SI{0}{\micro\siemens} with large tolerance error of \SI{15}{\micro\siemens} over approximately 10 iterations.
We also attempted to program the whole memory to \SI{75}{\micro\siemens} with a large tolerance error of \SI{15}{\micro\siemens}, but observed a dramatic decrease in the experimental F1 score. This can be attributed to the conductance difference distribution, which shows a clear peak around \SI{0}{\micro\siemens}. Although there is inherent noise, some devices are still precisely programmed to the same conductance, reducing the effectiveness of locality-sensitive hashing (see details in Supplementary Fig. 7). Compared to this method, resetting the crossbar array with several pulses is simpler to implement and also has lower energy consumption due to the lower conductance.

\subsection*{Balancing the `stay' errors and `skip' errors}
We use the same event detector as Scrappie \url{(https://github.com/ nanoporetech/scrappie)} to generate event vectors, which causes 50\% `stay' errors and 1\% `skip' errors\cite{work2_uncalled}. However, traditional content addressable memory can only search according to Hamming distance, which means the `stay' errors and `skip' errors need to be balanced. For the event vectors generated by Scrappie, 
$\mathbf{v}= [v_1, v_2, \ldots, v_n]$, we construct a new vector $\mathbf{u}$ such that an element $v_i$ is included in $\mathbf{u}$ if the absolute difference with its preceding element is greater than a predefined difference threshold 3 pA. Specifically, for each element $v_i$ from the second element of $\mathbf{v}$ (i.e., $i = 2$ to $n$), we include $v_i$ in $\mathbf{u}$ if $|v_i - v_{i-1}| > 3\ \text{pA}$. This process can be expressed as:
\begin{equation}
    \mathbf{u} = \{ v_i \mid |v_i - v_{i-1}| > 3, \, i = 2, 3, \ldots, n \}
\end{equation}
By filtering out the similar adjacent elements in $\mathbf{v}$, `stay' errors are significantly reduced, while `skip' errors may increase. In the scaled-up simulations, if a higher recall is desired, more possible difference thresholds can be tried like 4 pA.

Besides, a series of input seeds will be generated by selecting \textit{m} consecutive elements from the filtered events vector $\mathbf{u}$ to represent local information. The value of \textit{m} also needs to be well optimized; in practical applications, \textit{m} is set to ten to minimize the impact of the insertion and deletion (indel) errors while preserving uniqueness of the input seeds. To be more specific, because each event represents a 6-mer due to the nature of nanopore in R9.4 chemistry, a small \textit{m} results in excessive similar \textit{m}-events in the reference, compromising uniqueness. Conversely, a large \textit{m} results in indel issues introduced by `stay' errors and `skip' errors, making the input \textit{m}-events invalid as seeds for raw signal alignment.

Unlike traditional hashing and seeding algorithms, such as seed-and-vote\cite{liao2013subread} and seed-and-extend\cite{altschul1990basic} for nucleotide-based read mapping, which all require exact matches, and are susceptible to insertions and deletions in seeding. Here generate LSH seeds and perform approximate search with CAM in a high-dimensional space, using a predefined CAM threshold. Locality sensitive hashing can hash nearby regions in hyper-dimensional space to binary vectors with small Hamming distances. This means that even if sequencing errors occur, seeds are still likely to be close in the hyperplane. The parallel approximate search in CAM, using a predefined Hamming distance threshold, enables a more flexible and sensitive fuzzy seeding approach, matching seeds that are not perfectly identical but still very similar. This fuzzy seeds technique has also been reported to enhance sequence alignment sensitivity in genome assembly\cite{firtina2023blend} by identifying similar but not identical seeds.

\subsection*{Hardware-software co-design}
To fully utilize the intrinsic noise of memristor devices and parallel approximate search capability of CAM, we propose fuzzy seed-and-vote algorithm for memristor based raw signal alignment (\figref{fig2}). The parameter \textit{m} of events is set to ten, and four hundred rows of CAM arrays share a same location in SARS-CoV-2 read mapping scaled-up simulations, thus, one sub-array size is $400\times256$. We choose ten for \textit{m} to balance the uniqueness and errors, and four hundred rows for sub-array size depending on the reference genome size and max input samples. Only when the location with the maximum votes has at least twice as many votes as the second-highest location, and the maximum votes exceed a certain minimum threshold, we will consider the raw signal to belong to the reference genome of this location. When the locations with the highest and second-highest votes are adjacent, and their combined votes are more than twice those of the third-highest location, we consider the raw signal to belong to the reference genome between these two locations. We need to set a minimum votes threshold because nanopore sequencers sometimes produce very short sequence lengths. For example, if the maximum votes are three and the second-highest votes are one, they meet the ratio requirement, but three votes don't significantly exceed one vote, leading to many false positives if we still consider the raw signal to belong to the location with the maximum votes. We also show a detailed votes distribution example in Supplementary Fig. 8.

It's important to note that this method differs slightly from the experimental approach. In the proof-of-concept experiments, we used only one location for a single reference genome due to the limited number of memristor devices. Therefore, this experimental approach considers raw signals to be from the corresponding species when the votes for that location exceed a certain threshold in virus detection. However, in practical scenario, the input length is not fixed, so we cannot rely only on the maximum vote value to draw conclusions. Specifically, a 1,000 bp input will naturally receive more votes than a 400 bp input. If we set a fixed votes threshold based on the 400 bp input, the 1,000 bp reads will more easily surpass this threshold, increasing the risk of false positives.

\subsection*{Proof-of-concept experiments}
For the limited memristor devices, these proof-of-concept experiments have only one location for each species, corresponding to 78 bp randomly selected from the SARS-CoV-2 reference genome. In the virus detection experiment, we use the \texttt{--full-contigs} mode of Squigulator \cite{gamaarachchi2024simulation} to generate 1,000 MinION R9.4.1 raw signals in BLOW5\cite{gamaarachchi2022fast} format from the same 78 bp virus reference genomes. Then BLOW5 format raw signals are converted to FAST5 format using slow5tool\cite{samarakoon2023flexible} for further analysis. And the human raw signals are generated by Squigulator from random positions on human chromosome 1 with the same length of 78 bp. For the relative abundance estimation experiment, the raw signals are generated in the same way as virus raw signals.
We also compare the fully hardware implementation with simulations based on experimental results in the Supplementary Fig. 5. The results show that the F1 scores are very close, indicating that the simulation based on experiments can accurately represent the experiments for subsequent scaled-up simulations.

\subsection*{Scaled-up simulations}
To evaluate read mapping performance,  we randomly map 10,000 raw signals from a MinION R9.4.1 flow cell, sourced from the CADDE Centre, to the complete SARS-CoV-2 reference genome. The ground truth mapping positions are determined by Minimap2\cite{li2018minimap2} using the basecalled nucleotides in the dataset. If the positions identified by our system overlap with Minimap2's results, they are considered true positives. If they do not overlap, they are considered false positives. If our system does not identify a mapping position while Minimap2 does, it is considered a false negative. To compute F1 scores and other metrics, we integrate \texttt{UNCALLED pafstats} function\cite{work2_uncalled} in the scaled-up simulation code. The virus detection datasets are also real sequencing raw signals, while relative abundance estimation still uses simulated raw signals generated by Squigulator with default parameters.

\section*{Acknowledgements}
The authors would like to thank Sam Kovaka, Hasindu Gamaarachchi, and Can Firtina for their thorough explanations and insightful discussions.
This work was supported in part by RGC (7003-24Y, 27210321,  C1009-22GF, T45-701/22-R), NSFC (62122005), and ACCESS – an InnoHK center by ITC, Croucher Foundation, and Germany/Hong Kong Joint Research Scheme (GHKU707/23).

\section*{Competing interests}
The authors declare no competing interests.

\section*{Data availability}
The nanopore sequencing dataset and reference dataset are all publicly available. Source data for scaled-up simulation is available at \url{https://github.com/peiyihe/Mem_RT_Genomics}. Other data in this paper is also available with reasonable requests from corresponding authors.

\section*{Code availability}
The code for scaled-up simulations based on the real-world memristor conductance and real sequencing data is publicly available at the GitHub repository: \url{https://github.com/peiyihe/Mem_RT_Genomics}.

\bibliography{ref}

\end{document}